# Shared factory: a new production node for social manufacturing in the context of sharing economy

Pingyu Jiang [a,*], Pulin Li [b]

[a b] State Key Laboratory for Manufacturing Systems Engineering, Xi'an Jiaotong University, Xi'an, Shaanxi 710049, China

[*] Corresponding author. E-mail: pjiang@mail.xjtu.edu.cn, Tel: 029-83395396, Fax:029-83395153 Address: No.99 Yanxiang Rd. Xi'an, China. 710049

**ABSTRACT** Manufacturing industry is heading towards socialization, interconnection, and platformization. Motivated by the infiltration of sharing economy usage in manufacturing, this paper addresses a new factory model - shared factory, and provides a theoretical architecture and some actual cases for manufacturing sharing. Concepts related to three kinds of shared factories which deal respectively with sharing production-orders, manufacturing-resources and manufacturing-capabilities, are defined accordingly. These three kinds of shared factory modes can be used for building correspondent sharing manufacturing ecosystems. On the basis of sharing economic analysis, we identify feasible key enabled technologies for configuring and running a shared factory. At the same time, opportunities and challenges of enabling the shared factory are also analyzed in detail. In fact, shared factory, as a new production node, enhances the sharing nature of social manufacturing paradigm, fits the needs of light assets and gives us a new chance to use socialized manufacturing resources. It can be drawn that implementing a shared factory would reach a "win-win" way through production value-added transformation and social innovation.

## 1 Introduction

Social manufacturing is defined as a kind of internet-based and service-oriented manufacturing paradigm[1], in which manufacturing activities are going to more complex, diverse, and personalized[2], and factories are becoming more open, sharing, collaborative, and flexible[3]. Different from the traditional manufacturing paradigms, social manufacturing has its own distinctive characteristics, as Figure 1 shows.

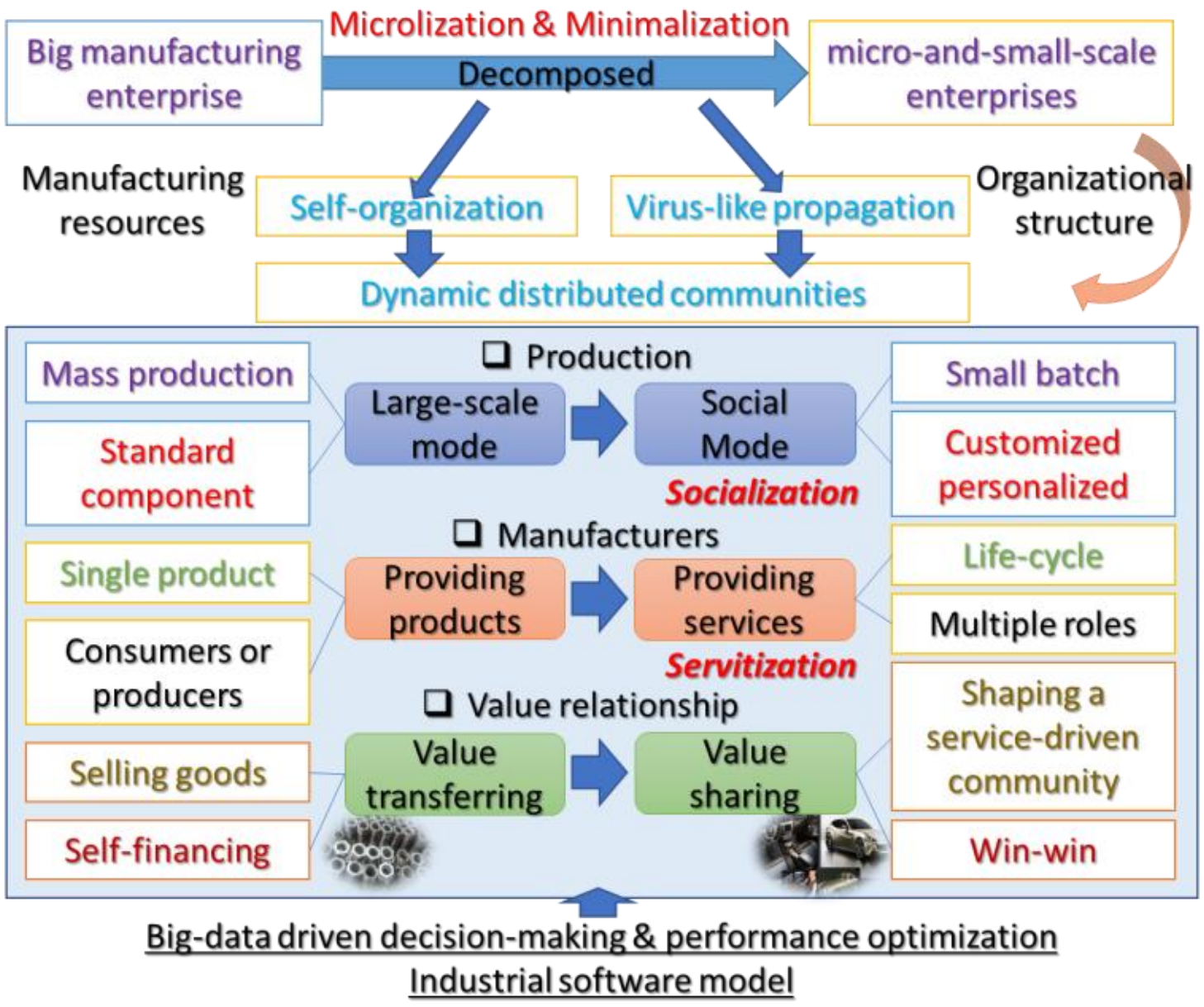

**Figure 1** Evolution of manufacturing paradigm and its characteristics

Here, the following characteristics need to be emphasized on in detail.

(1) Production changes from large-scale mode to customized and personalized ones in which batches are often smaller and smaller, even become one-of-a-kind[4]. Furthermore, the quality and delivery requirements of products are always customized. In addition, consumers are tending to focus on the reliability of products (usually endorsed by brands) instead of concerning on where and how the products manufactured.

(2) Manufacturers who run factories are providing services to customers instead of products, and these services might exist at any stage of product life-cycle[5]. In this way, the boundaries of manufacturers and customers are blurring with their roles' changes. It means that a factory might act as the role of consumer who uses manufacturing-capabilities from other manufacturers in one situation and be the role of producer who provide its own manufacturing-capability to other customers in another (called as prosumer role). This is an issue of service nesting. Here, manufacturing-capability sharing depends on manufacturers' playing a role of supplier in a service-driven mode. It must be pointed out that the customers are connected with either manufacturers mentioned above or pure users of manufacturing-capabilities who are actually without any manufacturing-resources attached to factories. So, manufacturing participants including both manufacturers and customers will pay more attention to their core business to achieve greater economic returns and present themselves with multiple roles.

(3) Value relationship are varying from value transferring to value sharing. In the beginning, manufacturers get profits by selling goods so they had to invest some fixed assets to complete the production-orders. Along with the role's changes, customers and manufacturers can shape a service-driven community of interests. In fact, such a community makes the value distribution more reasonable. At the same time, there is also a business coalition among similar manufacturers, which drives the business relationship from competition to collaboration for more mutual benefits.

Meanwhile, sharing economy, in which ordinary consumers also act as sellers, is attracting much interest[6] in a series of scenes such as transportation (Mobike[7], Uber[8]) and accommodation (Airbnb[9]). There is no doubt that the sharing economy has also penetrated into the manufacturing field, i.e., the sharing manufacturing. With the help of Information Technology (IT) and intermediary mechanism, manufacturers can share their productivity on the network or platform with relative low costs[10] and potential customers can also get enough market information and manufacturing-resources without limitation.

Distinguishing sharing manufacturing, which actually is a kind of implementation of social manufacturing, from other advanced manufacturing paradigms depends on the emergence of the platform at the very heart of the transaction. As shown in Figure 2, the platform occupies pride of place, brings manufacturers (who belong to different factories respectively and have various manufacturing-capabilities but without production-orders) and customers (who have production-orders but no manufacturing-capabilities) together, collects and disburses payments, and maintains the ratings-based cyber-credit system that makes the sharing marketplace work authoritatively. In a sense, such a kind of factory can be seen as a **Shared Factory**, in which ownership rights and usage rights are not same. Here, the platform actually is a front-end of the service-oriented community of interests mentioned above.

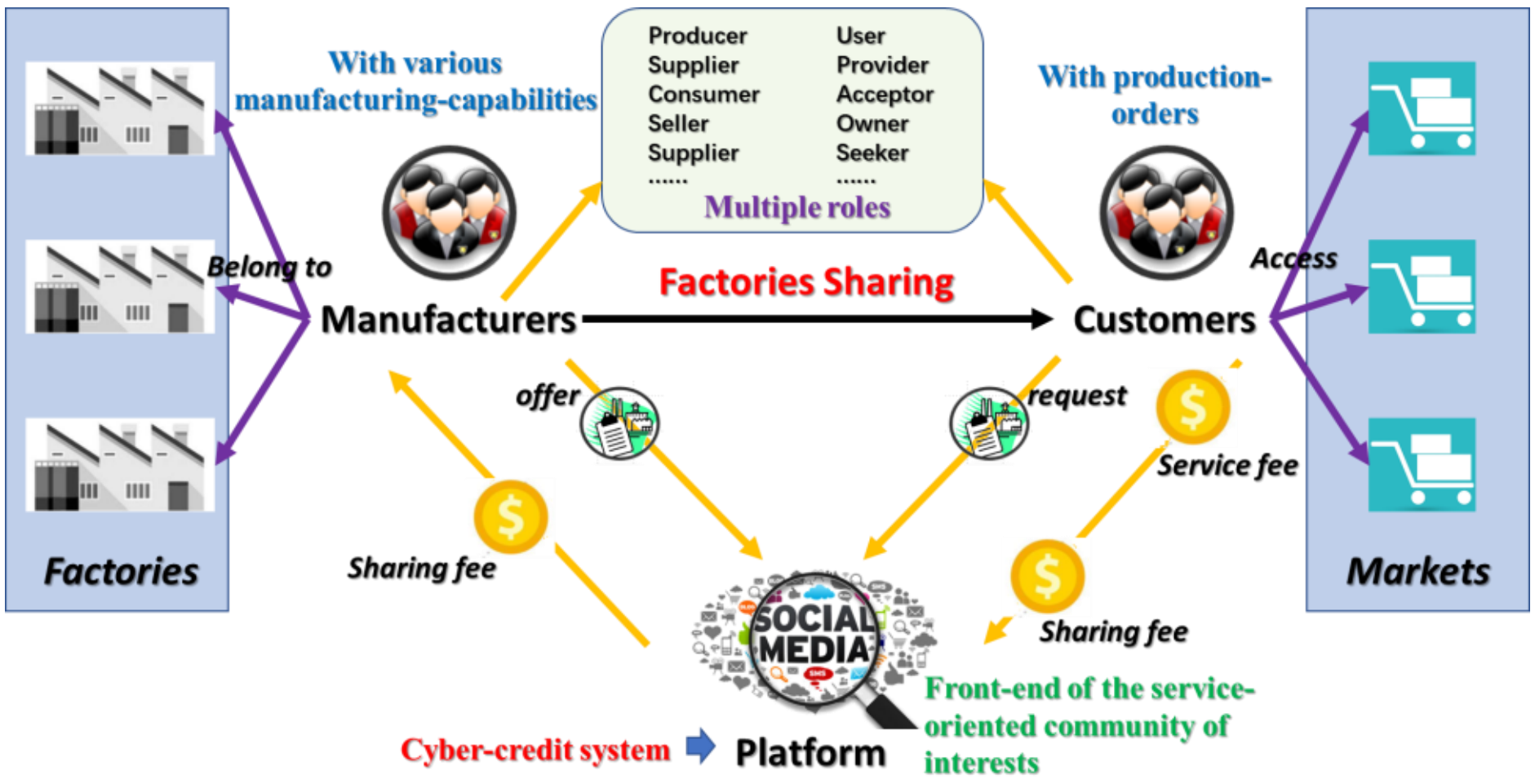

**Figure 2** Overview of the sharing relationship in manufacturing[11]

Although the shared factory model is new, the idea of sharing resources is not. For instance, traditional outsourcing can be seen as a kind of production-order sharing. There have been some practical and academic discussions of the sharing economy recently, but academic research on this topic, i.e., shared factory or sharing manufacturing is as quite limited yet. Some researchers have focused on the demand-allocation mechanisms driven by platforms [8, 12], and some others are concerning on how the sharing mechanisms affect the incumbent factories[13]. There are also papers that model a market where manufacturers can rent out their equipment to potential customers when not occupied, considering optimal pricing and product quality decisions.[14, 15]

In fact, shared factory is a new factory model in the context of sharing economy, works with a platform as its front-end of service-oriented community of interests, and is also a typical production node for social manufacturing. Concepts, characteristics, operation and interaction logic of the shared factory will be discussed in this paper. It would be helpful for us to not only understand the changes in manufacturing logic and the way of distributing benefits, but also figure out the value-added mechanism concerning such a shared factory model.

## 2 Clarification of shared factory

Shared factory is an extension of the sharing economy in the field of manufacturing. It is necessary to clarify the definition of the shared factory.

***Shared Factory*** is defined as a kind of factory-level manufacturing system with its fixed ownership, different rights for both different manufacturers from the inside and/or the outside, and different customers from the outside of the factory to use either the whole and partial manufacturing-resources, or manufacturing-capabilities in the manner of service and service-nesting mechanism so as to finish production-order-oriented manufacturing activities.

Manufacturer and customer take part in the manufacturing sharing by playing variety of roles in shared factory so that intertwined services form the service-nesting mechanism. Typical role set for both manufacturer and customer can be defined as

$$M_{role} := \{procumer, producer, sharer, factoryOwner, productionOrderAcceptor,$$
$$internalProducer, internalConsumer, \cdots\}$$

$$C_{role} := \{procumer, consumer, productionOrderProvider, mfgResourceConsumer,$$
$$capabilityConsumer, externalProducer, 3^{rd}\ ExternalProducer, \cdots\}$$

### 2.1 Classifications of shared factory

There are mainly three kinds of shared factories according to what the shared factory is used for:

(1) The FIRST kind of shared factory is used for sharing production-orders from others. Production-orders from either other manufacturers or customers might too many to deliver on time or have production-orders with specific needs, so these manufacturers or customers will share these production-orders to the shared factory so as to fill the gap of insufficient capabilities.

(2) The SECOND kind of shared factory is used for sharing its own manufacturing-resources to others. This kind of factory might own a lot of manufacturing-resources but doesn’t accept production-orders so as to share its own manufacturing-resources to either other manufacturers or customers who lacks sufficient manufacturing-resources to expand production.

(3) The THIRD kind of shared factory is used for sharing its own manufacturing-capabilities as services to either other manufacturers or customers on demand.

### 2.2 Characteristics of shared factory

Shared factory model has distinct characteristics compared to the traditional factory models, and is also quite different from some similar models, such as social factory[16] and trade association[17]. Distinctions among them are shown in Table 1.

**Table 1** Comparison of Shared factory mode and other similar ones

| | **Shared factory** | **Social factory** | **Trade Association** | **Traditional factory** |
|---|---|---|---|---|
| Property | Ownership rights and usage rights are different | Ownership rights and usage rights are same | | |
| Bargaining power | SMEs | Individual | Industry leader | Large & medium enterprises |
| Organization | Self-organizing & self-adjustment | | Alliance agreement | Industry gathering |
| Information architecture | Flat & distributed | | Centralized | Distributed |
| Sharing scope | Production-orders/ manufacturing-resources/ manufacturing-capabilities | Production-orders | Production-orders | N/A |
| Platform | Self-organizing platform / Third-party platform | | Industry platform | N/A |
| Resources effectiveness | Efficient | | Low | N/A |
| Openness | Fully open | Order open | Information open | N/A |
| Competition & cooperation | Cooperation dominant | | Internal cooperation, external competition | Competition dominant |

## 3 Shared factories in manufacturing ecosystems

Three kinds of shared factories are attached to three types of corresponding manufacturing ecosystems, i.e., production-orders-sharing-driven manufacturing ecosystem, resources-sharing-based manufacturing ecosystem, and capability-sharing-oriented manufacturing ecosystem. Their operational logics are shown in Figure 3, and some real-world cases of the shared factory can be used to illustrate these logics of the three types of sharing-driven manufacturing ecosystems.

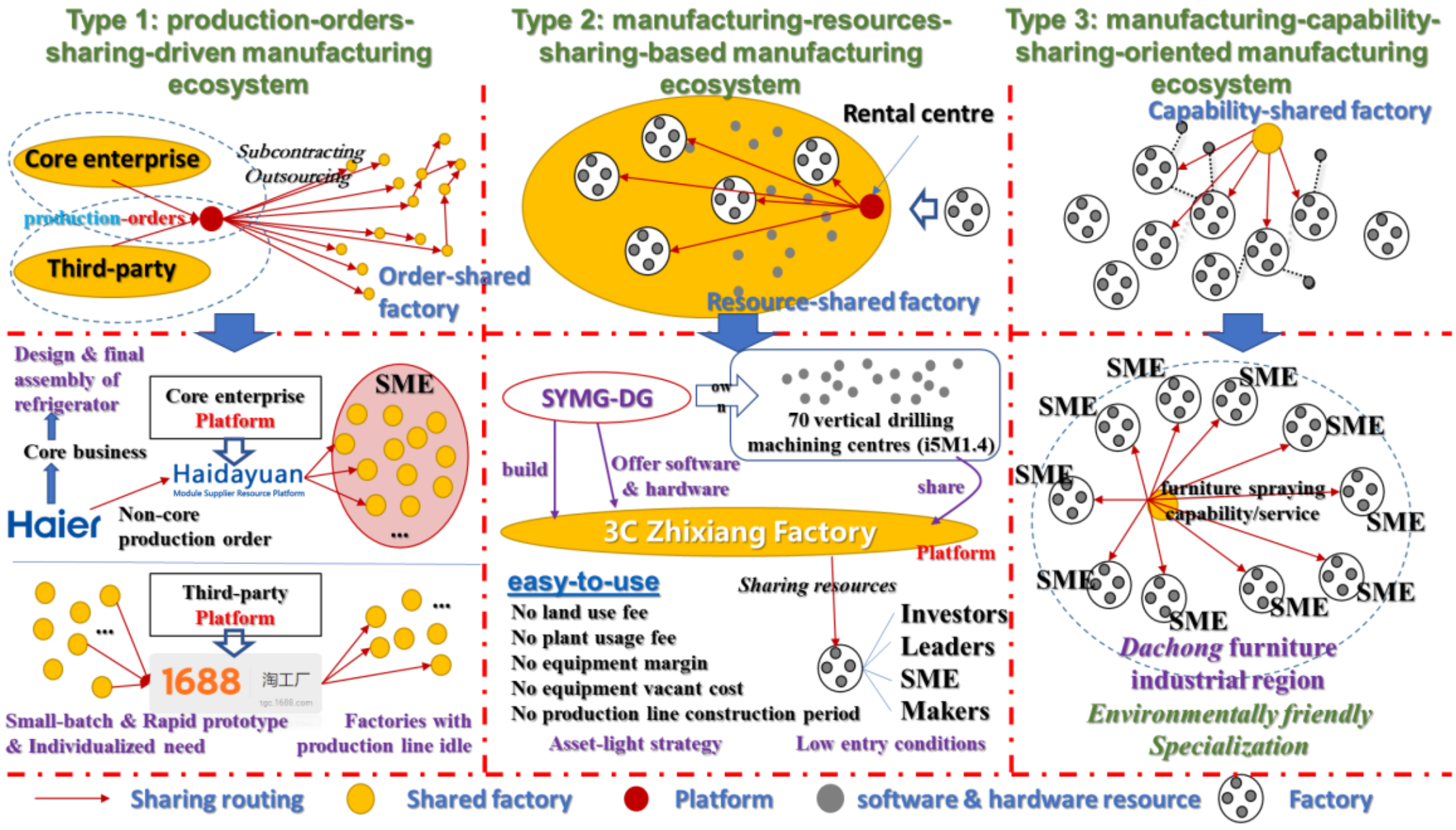

**Figure 3** The operational logics and cases of shared factories

The first type of sharing-driven manufacturing ecosystem depends mainly on *the first kind of shared factory* mentioned above. Where, shared factories can share production-orders from core enterprise platforms (such as *Haier Module Supplier Resource Platform - Haidayuan*[18]) or third-party platforms (such as *Alibaba's shared factory platform - 1688 Tao Factory*[19]). These platforms collect production-orders either from other factories whose production-orders are too many to deliver on time or with specific needs, or from customers who hold production-orders, are without manufacturing-resources and hope to share these shared factories for their manufacturing tasks. In this way, core enterprise or third-party platform can actively coordinate the available manufacturing-resources in the ecosystem, so that the production-order could be completed as soon as possible. Shared factory in this type of manufacturing ecosystem can be considered as an extension of traditional outsourcing and crowdsourcing nodes.

The second type of sharing-driven manufacturing ecosystem deals with *the second kind of shared factory* mentioned above. Where, shared factories act as the platform role through which they share their own hardware and software resources (including plant, machines, workers and so on) out, such as the *3C Zhixiang Factory*[20] built by *Shenyang Machine Tools (Dongguan) Intelligent Equipment Co., Ltd. (SYMG-DG)* in Guangdong Province, China. They will integrate and manage their own idle manufacturing-resources, and share these resources to customers (usually SMEs) who need. In fact, customers can bring their own production-orders into a shared factory, use the suitable manufacturing-resources of the shared factory to complete production-orders, and charge a certain fee to the shared factory according to the resource usage time or some other criteria.

The third type of sharing-driven manufacturing ecosystem is concerned with *the third kind of shared factory* mentioned above. Where, shared factories have specific manufacturing-capabilities and can share these capabilities as services to customers (usually SMEs) who don’t want to invest and build new factories. For instance, in *Dachong* town of Zhongshan city, Guangzhou province, there are a lot of similar companies engaged in the production of mahogany furniture[21]. The furniture spraying process always brings a lot of waste gas, so some shared factories with perfect paint exhaust gas treatment systems have emerged. They provide a complete spraying service for a lot of *Dachong* furniture factories without exhaust emissions. Especially in an industrial region or park, this sharing mechanism that relies on professional services powered by the shared factories is needed urgently.

It can be said that shared contexts will be changed if type of shared factories changes. For example, *3C Zhixiang Factory* is usually a manufacturing-resource-sharing-based shared factory when it only shares manufacturing-resources for customers. But it can also be changed into a manufacturing-capability-sharing-oriented shared factory when it uses its manufacturing-capabilities as services for designing, machining and assembling for its customers. This means that the type of a shared factory can be changer from one to another, and any a manufacturing ecosystem can use different kinds of shared factories and their combination to power its functions and performances.

## 4 Economic analysis of shared factory model

Shared factories, reintegrating the socialized manufacturing-resources, are a category of shared economy essentially, so the biggest benefit of this model is in terms of economics. A production function relates quantities of output of a production process to quantities of inputs, and refers to the expression of the technological relation between manufacturing-resources and production-orders as inputs, and goods as outputs.[22] It seems difficult for traditional factories to balance their manufacturing-resources and production-orders. Sometimes, factories have sufficient manufacturing-resources but few production-orders, while others have sufficient production-orders but lack manufacturing-resources. The emergence of shared factories is to resolve the contradiction between these two constraints.

Manufacturers and customers, no matter in which type of shared factory, have production-orders or manufacturing-resources, or both. The sharing process achieves the recombination of their redundant production-orders and manufacturing-resources. Cobb-Douglas production function[23] is a particular functional form of the production function, widely used to represent the technological relationship between the amounts of two inputs. Thus, the economic analysis of the shared factory model can be illustrated in Figure 4 with Cobb-Douglas production function, in which two inputs are the production-orders and manufacturing-resources.

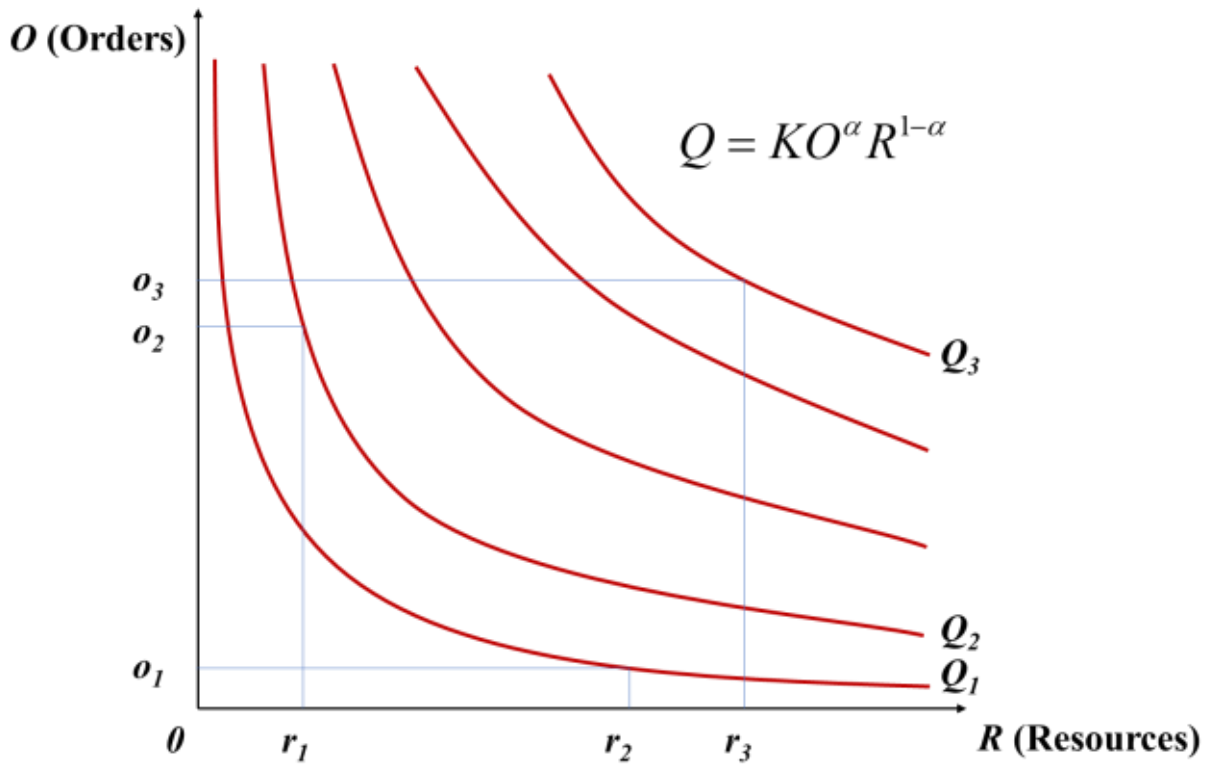

**Figure 4** Cobb-Douglas production function of the shared factory

The total production $Q$ can be expressed as

$$Q = KO^{\alpha}R^{1-\alpha}$$

where:

- $K$ - total factor productivity
- $O$ - amount of production-orders owned
- $R$ - manufacturing-resources invested
- $\alpha$ - output elasticities of production-orders

In Figure 4, production-order quantities of manufacturer and customer are $o_1$, $o_2$, and the manufacturing-resource inputs are $r_1$, $r_2$ respectively. Supposed that $o_3 = o_1 + o_2$, $r_3 = r_1 + r_2$, then

$$Q_3 = Ko_3{}^{\alpha}r_3{}^{1-\alpha} = K\left(o_1 + o_2\right)^{\alpha}\left(r_1 + r_2\right)^{1-\alpha} = K\left(o_1{}^{\alpha}r_1{}^{1-\alpha} + o_2{}^{\alpha}r_2{}^{1-\alpha} + R_n\right) = Q_1 + Q_2 + KR_n$$

since $K > 0$, $R_n > 0$, so $Q_3 > Q_1 + Q_2$, which means when the manufacturer and customer share their production-orders and manufacturing-resources with each other, they will get more production in total from overall view. And when the market is in a downturn ($Q$ is reduced), the shared factory model can also adjust the $O$ and $R$ of each participant simultaneously, so that each of them can maintain production without closure.

## 5 Identification of key enabled technologies for shared factory

Running a shared factory needs to get the support of key enabled technologies. So, it is quite important to identify what are key enabled technologies for shared factory. In general, there are six main supporting technologies to ensure the smooth operations of a shared factory, as shown in Figure 5.

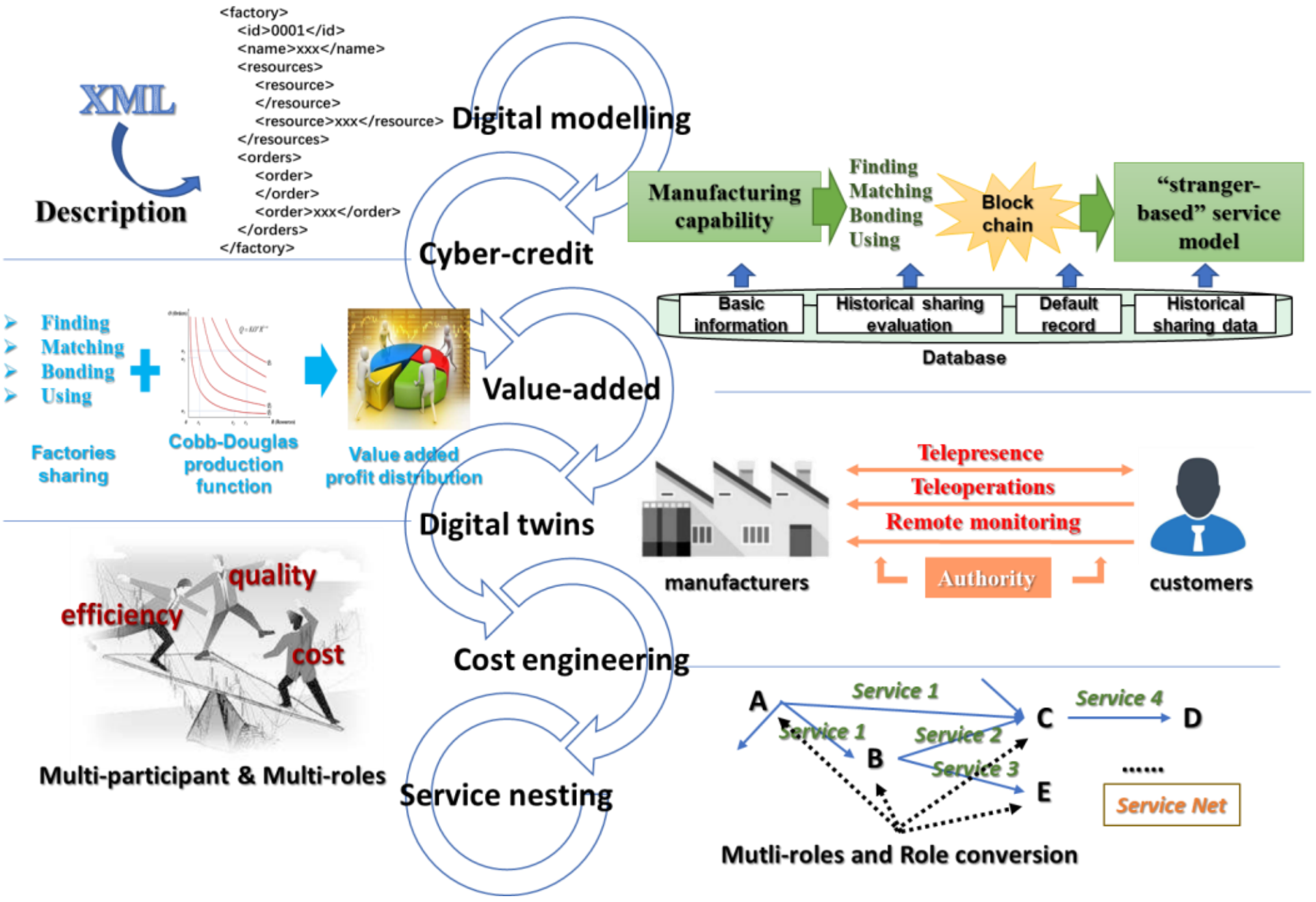

**Figure 5** Six key enabled technologies support the operations of a shared factory

### 5.1 Digital modelling of shared factory and its running mechanism

As mentioned above, two main factors involved in the shared factory are the production-orders and manufacturing-resources and evolve the third factor, that is, manufacturing-capabilities, which influences how to run the factory. To facilitate digital modelling of shared factories' operations, the *eXtensible Markup Language* (XML) schema[24] is introduced to clarify the relationship among participants including both manufacturers and customers, manufacturing-resources, production-orders, and manufacturing capabilities. It can not only show the basic properties of an actual factory (e.g., ownership), but also see who is using and operating these resources. Digital modelling of shared factory and its running mechanism is the fundamental of the five following enabled technologies.

### 5.2 Manufacturing-capability finding, matching, bonding and using mechanisms under "stranger-based" service and cyber-credit model

It is necessary to set up manufacturing-capability finding, matching, bonding and using mechanisms for the socialized manufacturing resources, while participants sharing the shared factory might be a kind of stranger as its collaborators. These mechanisms will enable efficient connections of manufacturing-capabilities and requirements, production-orders and services, and factories and manufacturing-resources. Noted that one of the most challenges is that all the above mechanisms need to be built on "stranger-based" service model. At the same time, manufacturing cyber-credit model is a codified compulsory, deterrent, and incentive mechanism governed autonomously by a smart contract system based on the distributed blockchain technology.[25] In a shared factory, cyber-credit model acts as the trusted third party to prevent defaults and frauds during manufacturing sharing interactions, and bridges the credit gap between manufacturers and customers.

### 5.3 Value-added mechanism based on sharing manufacturing-capabilities

Section 4 has concerned on the economic value of shared factory from the view of overall. However, the relationship between resources input and profits output/distribute of each participant still needs further discussion. Although shared factory has proven to be a "win-win" model, how the increased profits distribute reasonably is the key to determining whether this factory model is continuable and sustainable. Sharing a factory can enlarge manufacturers and customers bargain power and common profits, while each participant who join such a sharing can win its profit according to its contribution on either production-orders, manufacturing-resources or manufacturing-capabilities. For customers, sharing a factory provides a new value-added way through using additional manufacturing-resources and manufacturing-capabilities to get more benefits. For manufacturers who hold the shared factories, sharing a factory can also make the value added by getting into larger markets with taking more production-orders, and contributing their manufacturing-resources and manufacturing-capabilities to customers.

### 5.4 Digital twins as a telepresence, teleoperations, or remote monitoring under the control of authority

It must be emphasized that shared factory model makes it possible to closely connect the manufacturers with customers. They create sharing relationships dynamically through the above mechanisms although the physical distance between them is probably very far sometimes. So, it is necessary to use a digital twin model for realizing the telepresence, teleoperations, or remote monitoring of the manufacturing processes under the control of authority. Here, digital twin technology makes physical devices and virtual software connected seamlessly. Through creating a digital twin model related to operations, for example, real-time monitoring data in the physical world can be updated to the virtual model simultaneously and the controlling commands from the virtual model would be transmitted to the physical devices. In this way, it is possible for customers to be able to operate the shared factory especially when a manufacturing-resource-sharing-based shared factory runs.

### 5.5 Cost engineering related to three kinds of shared factory models

Cost engineering[26] plays an important role in taking a balance between cost, quality and efficiency. It deals with activity-based quality and efficiency estimation, cost control and forecast, investment appraisal and risk analysis. Due to the multiple participating roles in the shared factory model, cost engineering in this context needs to achieve the joint cost management, comprehensive efficiency assessment and collaborative quality control. It is convinced that cost engineering will enable the credible analysis of sharing alternatives, accurate quantification of the sharing capability, comprehensive balance in cost, quality and efficiency related to all sharing participants, *etc*. One of the typical methods for cost engineering relies on activity-based costing model.

### 5.6 Production-order planning, scheduling and service nesting

Generally, a production-order can be decomposed into several sub-production-orders according to the product bill of materials (BOM). A customer, who holds this production-order and acts as "*producer*" role, either share these sub-production-orders with production-order-driven shared factories, or uses manufacturing-resources or manufacturing-capability services form potential manufacturers who run respectively in the form of other two kinds of shared factories. In fact, this customer has two roles, that is, "*producer*" role when offering production-order-based products and "*consumer*" role when sharing three kinds of shared factories. This situation can be seen as a two-layer service nesting, which deals with sharing shared factories and realizing product providing services. In actual production, the service nesting is complicated by different roles of manufacturers and customers. In addition, other service providers such as cutting-tool service system can support production activities inside a shared factory. It is also a kind of service nesting. Therefore, it is necessary to explore the service nesting mechanism when the customer plans and schedules the correspondent production-order and sub-production-orders, and the manufacturers want to cut down their production cost through introducing some service providers.

## 6 Discussion

### 6.1 Opportunities

The shared factory model as a kind of new production node inside a social manufacturing network enhances the nature of manufacturing sharing in three forms which deal respectively with sharing production-orders, manufacturing-resources and manufacturing-capabilities. Its working logic relies on the platform which integrates different shared factories under the context of social manufacturing paradigm and is the front-end of a manufacturing community. This gives us a new way to use socialized manufacturing resources under the guidance of sharing economy. Both manufacturers and customers, who can take different roles concurrently under the context of running a shared factory, will share the benefits in a "win-win" mode. It is possible for any traditional factories including SMEs to use this shared factory model to broaden the market, optimize resource allocation, enhance the bargaining power, cut down the manufacturing costs, and increase the flexibility of production, together with the creative vitality inspiring.

### 6.1 Challenges

The shared factory model will be used mainly in a "*stranger-based*" service mechanism. In addition, remote operations related especially to a manufacturing-resource-sharing-based shared factory in terms of using digital twins require the very strict conditions of security. This big challenge makes the implementation of shared factory model become more complicated. One of the most feasible methods is to introduce new cyber-credit model such as block chain so as to ensure the security mentioned above. The other big challenges include how to develop an efficient costing model and build a value-added mechanism so as to let potential manufacturers and customers believe that "win-win" mode works well.

## 7 Conclusion

Shared factory, as a kind of new factory model with its fixed ownership, different rights and different customers, is becoming a new production node for social manufacturing in the context of sharing economy. It fits the needs of light assets and give us a new chance to use socialized manufacturing resources.

Shared factory model enhances the sharing nature of social manufacturing paradigm in detail. Such a sharing nature is connected with three implementing levels, that is, sharing production-orders, sharing manufacturing-resources and sharing manufacturing-capabilities. Here, platform would play an important role. It is very clear that this model also shows some contributions to production value-added transformation and social innovation. However, there are still some problems to be solved, such as how to use new IT technology in building a more feature-rich platform to integrate different shared factories, how to design a more rational profit distribution model, *etc*.

The future work will include developing the common implementation architecture of shared factory platform based new IT technology and exploring more enabled technologies for more interactive and collaborative manufacturing sharing.

### Acknowledgments

The research work is under the financial supports of both MOST's innovation method working projects with grant No. 2016IM010100 and natural science foundation of China with grant No. 71571142.

### Compliance with ethics guidelines

Pingyu Jiang and Pulin Li declare that they have no conflict of interest or financial conflicts to disclose.